\documentclass[11pt]{article}
\usepackage{epsfig}
\usepackage{amsfonts}
\usepackage{rotating}
\usepackage{cite}

\newcommand{\beq}{\begin{equation}}
\newcommand{\eeq}{\end{equation}  }
\newcommand{\la}{\langle}
\newcommand{\ra}{\rangle}
\newcommand{\bec}{\begin{center}}
\newcommand{\eec}{\end{center}}

\def\bgf{\mathrm{BGF}}
\def\pa{\uparrow\uparrow}
\def\an{\uparrow\downarrow}

\def\nc{n_{\mathrm{c}}}
\def\nt{n_{\mathrm{t}}}

\def\sBGFp{\sigma_{\mathrm{BGF}}^{\uparrow\uparrow}}
\def\sBGFu{\sigma_{\mathrm{BGF}}^{\uparrow\downarrow}}
\def\stotp{\sigma_{tot}^{\uparrow\uparrow}}
\def\stotu{\sigma_{tot}^{\uparrow\downarrow}}

\def\Nc{\tilde{n}_{\mathrm{c}}}
\def\Nt{\tilde{n}_{\mathrm{t}}}

\def\p+p{\pi^{\pm}\mbox{p}}
\def\K+p{\mbox{K}^{+}\mbox{p}} 
\def\m+p{\mu^{+}\mbox{p}}

\begin{document}

 
\hfill ANL-HEP-PR-99-89 
 
\hfill August  1999

\begin{center}
 

{\Large\bf Current-Target Correlations as a Probe of  
$\Delta G/G$ \\ in Polarized Deep Inelastic Scattering \\[-1cm]} 

\vspace{2.0cm}

{\large 
I.~V.~Akushevich $^a$ and  S.~V.~Chekanov $^b$\footnote{ 
On leave from
the Institute of Physics,  AS of Belarus,
Skaryna av.70, Minsk 220072, Belarus.}}

{\begin{itemize}
\itemsep=-1mm
 
\normalsize
\item[$^a$]
\small
National Center of  Particle and  High Energy Physics,
Bogdanovich str. 153, 220040 Minsk, Belarus. \\
Email: akush@hermes.desy.de

\normalsize
\item[$^b$]
\small
Argonne National Laboratory,
9700 S.Cass Avenue,
Argonne, IL 60439. 
USA \\ Email: chekanov@mail.desy.de 

\end{itemize}
}

\normalsize

\vspace{1.0cm}
\begin{abstract}
The measurement of the polarized gluon distribution function
$\Delta G/G$ using current-target correlations
in the Breit frame of deep inelastic scattering is proposed.
The approach is illustrated using a Monte Carlo simulation of 
polarized $ep$-collisions for HERA energies.  
\end{abstract}


\end{center}

\vspace{1.0cm}


\section{Introduction}
A direct determination of the polarized gluon distribution is of
importance for understanding  the spin properties of the nucleon. 
Recent experimental measurements of spin-dependent structure 
functions \cite{CERN,SLAC,DESY}
have shown that the valence quarks account for only a small fraction of
the nucleon spin. One of the possible 
explanations for this observation,  
known  as ``spin crisis'',  is to assume a  
large contribution from the gluon spin.  
There exist many theoretical models 
which explain this phenomenon, but there is no experimental
evidence to favor one of them.

A direct way to solve this puzzle is to measure 
the gluon density in polarized lepton-nucleon scattering. 
One of the suggested methods is based on the detection of the 
correlated high-$p_t$ hadron pairs in polarized deep inelastic
scattering (DIS) \cite{brav}.
This measurement has already been performed  by the 
HERMES Collaboration \cite{Amarian}. 
Another possibility, which is planned to be studied at the COMPASS experiment,
is to analyze  
events with open/closed charm \cite{COMPASS}. 
At the HERA $ep$ collider, after possible upgrade to polarized beams,
the polarized gluon density can also be measured
from dijet events \cite{roeck}.  

\begin{figure}
\begin{center}
\mbox{\epsfig{file=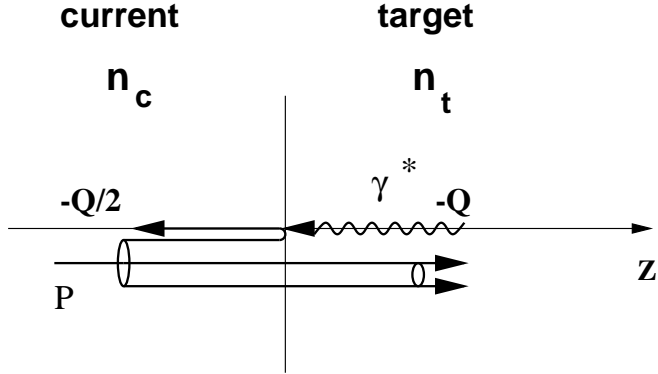,height=5.0cm}}
\caption{ {\it
A schematic representation of  the Breit frame for the
quark-parton model. }}
\label{fig1}
\end{center}
\end{figure}

In this paper a  possibility to measure $\Delta G/G$
from the multiplicity correlations in neutral-current DIS is discussed.
The proposed  method is based  on the Breit frame \cite{Br}.
For this frame, in the quark-parton model (QPM),
the incident quark carries $Q/2$ momentum\footnote{As usually,
$Q^2=-q^2=-(k-k^{'})^2$ ($k$ and $k^{'}$ denote the 4-momenta of the
initial and final-state lepton, respectively)
and the Bjorken
scaling variable $x$ is defined as $Q^2/(2 P\cdot  q)$,
where $P$ is the
4-momentum of the proton.} 
in the positive $Z$-direction and the outgoing
struck quark carries the same momentum in the
negative $Z$-direction. 
The phase space of the event can be divided into two
regions (see Fig.~\ref{fig1}).
All particles with negative $p_Z^{\mathrm{Breit}}$ components
of momenta form the current region. In the QPM,
all these particles are produced
from  hadronization  of
the struck quark. Particles with positive $p_Z^{\mathrm{Breit}}$  are
assigned to the target region, which is associated with the
proton remnants.

For the QPM in the Breit frame,
the phase-space separation between  the particles 
from the struck quark and the proton remnants is maximal. 
Therefore, it is
expected that there are no correlations between   
the current-region particles and particles in the target region of the
Breit frame \cite{ch}.
Such a separation does  not exist when
the leading-order QCD processes, known as Boson-Gluon Fusion (BGF) 
and QCD-Compton (QCDC) scattering, are involved \cite{str}. 
The kinematics of these processes lead to current-target anti-correlations
predicted in \cite{ch} and experimentally measured by the ZEUS 
Collaboration \cite{ZEUS}. The magnitude of these 
correlations at small $x$ is mainly
determined by the BGF processes.
 
\section{Current-Target Correlations. Unpolarized Case}

Recently, it was noticed that the 
BGF events in non-polarized DIS can be studied  without
involving  the jet algorithms \cite{ch}.  For this one
can measure a liner interdependence between the
current- and target-region multiplicities  in  the Breit frame.
The approach involves the measurement of the particle
multiplicities  in  large phase-space regions, rather than
clustering separate particles at small resolution scales. 
Therefore,  high-order QCD 
and hadronization effects are expected to be smaller than
for the dijet studies that use clustering algorithms with 
specific resolution scales.  
In addition, the method does not depend on  
the jet transverse energy $E_T$ used in jet reconstruction and 
is well suited
for low $Q^2$ regions where the jet algorithms are less reliable.

The correlation  between the current and target region
multiplicities  can be measured with the covariance
\beq
\mathrm{cov}= \la \nc\, \nt \ra -
\la \nc \ra \la \nt \ra,
\label{1}
\eeq
where $\nc$ ($\nt$) is  the number of final-state particles in the
current (target) region and $\la \ldots \ra$ is the 
averaging over all events.
If $h$  particles from the hard QCD processes 
are emitted  in the target region, one can rewrite
(\ref{1}) as
\beq
\mathrm{cov}  =
\la (\Nc -h)\, (\Nt +h) \ra - 
\la \Nc -h \ra \la \Nt +h \ra .
\label{2}
\eeq
Here  $\Nc$ is the total number of particles coming from    
zero  and first-order QCD processes ($h\le \Nc$) and 
$\Nt$ is the  multiplicity of the proton remnants  without
counting the particles  due to the hard QCD  processes.  
From (\ref{1}) one obtains
\beq
\mathrm{cov} = \la \Nc\, h \ra - 
\la \Nc \ra  \la h  \ra  -  \la h^2  \ra + \la h  \ra^2 . 
\label{3}
\eeq
In this expression, the contribution from the remnant
multiplicity $\Nt$ cancels  
since we consider the case when $\Nt$ is  independent
of  $\Nc$ and $h$. This key assumption means that the only important  
effects leading to the correlation between $\nc$ and $\nt$ are
the hard QCD radiation.  
The validity of this assumption has 
been  tested  in \cite{ch} for a Monte Carlo
model based on the parton shower, describing higher than first order
QCD effects, and the LUND string model for hadronization. 

At small $x$, the BGF is the only dominant first-order QCD  process.
Let us define the BGF  rate as 
\beq
R_\bgf=\frac{N_{\bgf}}{N_{\mathrm{ev}}}, 
\label{4}
\eeq
where $N_\bgf$ is the number of  the BGF  events
and $N_{\mathrm{ev}}$ is the total number of events 
in the limit $N_{\mathrm{ev}}\to\infty$.
According to the assumption discussed above,  $h=0$ for the QPM. 
For the BGF, however, one or  two quarks are emitted in the target 
hemisphere, so that their parton radiation gives $h>0$. 
The averaging in (\ref{1}-\ref{3}) is performed  over all
possible DIS events, despite the fact that  $\mathrm{cov}\ne 0$ 
for the BGF events only. Therefore,  one can apply  
the averaging over the relevant BGF events using 
the relation:   
\beq
\la f(h) \ra = R_\bgf \> \la f(h) \ra_\bgf ,  
\label{5}
\eeq
where $f(h)$ is any function of $h$ so that $f(h=0)=0$ and
$\la\ldots \ra_\bgf$ is the average over the BGF events.  
Using this relation, one has 
\beq
\mathrm{cov} = \left(\la \Nc\, h \ra_\bgf -
\la \Nc \ra  \la h  \ra_\bgf  -  
\la h^2  \ra_\bgf + R_\bgf \la h  \ra^2_\bgf\right)\> R_\bgf.
\label{6}
\eeq
The last term can be neglected since it contains  a 
small contribution.
Therefore, in the linear approximation, 
the covariance  can be expressed as  
\beq
\mathrm{cov} \simeq  - A(Q^2,x) \>  R_\bgf     
\label{7}
\eeq
with    
\beq
A(Q^2,x) = \la \Nc \ra  \la h  \ra_\bgf + 
\la h^2  \ra_\bgf - \la \Nc\, h \ra_\bgf . 
\label{8}
\eeq  
A more detailed form of 
this expression has been obtained in \cite{ch}.

The function $A(Q^2,x)$ depends on:  

1) A number of the final-state hadrons in the jets
initiated by the quarks in the BGF processes. 

2) Kinematics of the BGF jets in the Breit frame, 
which depend on $Q^2$ and  $x$. Using
LEPTO Monte Carlo \cite{lepto} with the tuning described in \cite{tun}
and  GRV94  HO  \cite{grv}
parameterization of the parton distribution functions, 
we have found  that for  $5 < Q^2 <50$ GeV$^2$,
the fraction of BGF events with both quarks moving to the target
region increases  from $52\%$ at $\la x\ra \simeq 0.2\cdot 10^{-1}$ 
to $61\%$ at $\la x \ra =0.5\cdot 10^{-3}$
($100\%$ corresponds to all possible jet configurations of the BGF events in
the Breit frame).
Therefore, $A(Q^2,x)\simeq A(Q^2)$ in (\ref{7})  is a good approximation.
Note an $x$-dependence of the $\mathrm{cov}$ due to the BGF kinematics
is rather small compared to the $x$-dependence of
the BGF rate itself:  $R_\bgf$
increases  from $7\%$ at $\la x\ra \simeq 0.2\cdot 10^{-1}$ to 
$21\%$ at $\la x \ra =0.5\cdot 10^{-3}$. 
 
Before going to a polarized case, 
let us again remind all approximations
made in (\ref{7}).

\begin{enumerate}
\item
We neglect the QCD Compton  scattering, considering low 
$Q^2$ regions.  Note that even  if a small fraction of the QCDC
is present, some QCDC events cannot contribute to the correlations since
singularities in the QCDC cross section favor the event topology with two
jets in the current region, which does not produce the correlations
(see, for example, Ref.~\cite{chek2}). 
 
\item
Effects from high-order QCD and hadronization
are already assumed to contribute to a  value of $A(Q^2,x)$, an exact form
of which is beyond the scope of the present study. However,
the assumption made to derive (\ref{7}) was that
the high-order QCD processes and hadronization do not change significantly
the LO dijet kinematics in the Breit frame. This
assumption leads to the factorization of $A(Q^2,x)$ from the BGF rate
in the linear approximation.         
\end{enumerate} 

Using a numerical simulation \cite{ch}, it was shown that the effects
quoted above do not strongly contribute to the 
relation (\ref{7}) with an $x$-independent $A(Q^2)$  in the range
$5<Q^2< 50$ GeV$^2$ and $0.5\cdot 10^{-3} < x < 0.2\cdot 10^{-1}$.
According to  the LUND model implemented in JETSET,
the hadronization  does not produce large
correlations during the formation and independent 
breaking of the strings
stretched between the  current-region
partons and the remnant.

\section{Polarized Case}

The ratio of spin dependent gluon density 
$\Delta G(x)$ to  spin averaged gluon density $G(x)$ at a fixed $Q^2$
is proportional to the asymmetry of the BGF cross sections,
\beq
\la a_{LL}\ra\>  \frac{\Delta G(x)}{G(x)} = 
\frac{\sBGFu - \sBGFp}{\sBGFu + \sBGFp}=
\frac{\stotu\> R_\bgf^{\an} - \stotp\> R_\bgf^{\pa}} 
{\stotu\> R_\bgf^{\an} + \stotp\>  R_\bgf^{\pa}}, 
\label{12}
\eeq
where $\la a_{LL}\ra$ is the value of 
the BGF asymmetry at partonic level \cite{poldis}, 
$\sBGFu$ ($R_\bgf^{\an}$) and $\sBGFp$ ($R_\bgf^{\pa}$) 
are  the BGF cross sections (BGF rates)  
in case of the ant-parallel  ($\an$) and parallel ($\pa$)
polarizations of the incoming lepton
and nucleon. $\stotu$ and $\stotp$ are the total cross sections
which can be obtained by counting the number of events in a 
given kinematic bin  for the  
different spin configurations, normalized to an integrated luminosity.

The expression (\ref{12}) can further be rewritten as  
\beq
\la a_{LL}\ra\> \frac{\Delta G(x)}{G(x)} \simeq \Delta\rho + A_{||}, \qquad
\Delta\rho=\frac{R_\bgf^{\an} - R_\bgf^{\pa}}{R_\bgf^{\an} + R_\bgf^{\pa}},   
\label{13}
\eeq
\beq
A_{||}=\frac{\stotu - \stotp }{\stotu + \stotp} ,  
\label{14}
\eeq
where $A_{||}$  
is an  inclusive polarized asymmetry, which is
expected to be small compared to the asymmetry $\Delta\rho$ of  
the BGF rates. The relationship between (\ref{12}) and (\ref{13}) can
easily be seen after rewriting $\Delta\rho + A_{||}$ as
$$
\frac{\stotu\> R_\bgf^{\an} - \stotp\> R_\bgf^{\pa}}
{(\stotu\> R_\bgf^{\an} + \stotp\>  R_\bgf^{\pa})(1-a)},
\qquad
a=\frac{(R_\bgf^{\an} - R_\bgf^{\pa})(\stotu-\stotp)}
{2(\stotu\> R_\bgf^{\an} + \stotp\> R_\bgf^{\pa})},   
$$
which is approximately equal to $(\ref{12})$ by neglecting a small  
contribution from $a$. 

According to (\ref{7}),  
\beq
\mathrm{cov}^{\an}\simeq -A(Q^2)\> R_\bgf^{\an},  
\qquad
\mathrm{cov}^{\pa}\simeq -A(Q^2)\> R_\bgf^{\pa},  
\label{15} 
\eeq
where $A(Q^2)$ is considered to be independent of  the polarization, 
since this function is mainly determined by  
the number of the final-state hadrons in jets, i.e. 
by multiple-gluon radiation and hadronization.  
From (\ref{15})  
it is easy to see that the theoretical $\Delta\rho$ 
can be determined via the covariances  
\beq
\Delta\rho\simeq
\Delta\mathrm{cov} = \frac{\mathrm{cov}^{\an} - \mathrm{cov}^{\pa}}
{\mathrm{cov}^{\an} + \mathrm{cov}^{\pa}}.  
\label{16}
\eeq
Thus the asymmetry $\Delta G(x)/G(x)$ can experimentally be  obtained  
from $\Delta\mathrm{cov}$ and directly measurable $A_{||}$. 
Below we shall numerically estimate the BGF asymmetry 
$\Delta\rho$ (\ref{13}). We calculate this quantity directly
from the BGF rates  and then reconstruct it 
by measuring the $\Delta\mathrm{cov}$ from the final-state hadrons
of polarized DIS.      

\section{Numerical Studies}

In order to study the asymmetry using the current-target correlations
we use the PEPSI Monte Carlo generator \cite{PEPSI} for the polarized 
leptoproduction. 
This model is based on the LEPTO 6.5 for DIS together with
JETSET 7.4 describing the fragmentation. Two DIS samples with the opposite
spin configurations were generated in the region $5<Q^2< 50$ 
GeV$^2$. The electron and proton beam energies 
were taken to be 27.5 and 920
GeV, respectively. For simplicity, 
both beam polarizations were assumed to be $100\%$. 
The total number of the generated DIS events is 1M for
each polarization sample for the given $Q^2$ range. 
The covariances for each polarization  were
determined from charged final-state hadrons according to (\ref{1}).
Hadrons with lifetime $c\tau >1$ cm  are declared to be stable.

\begin{figure}
\begin{center}
\begin{picture}(180,220)
\put(-100,0){\epsfig{file=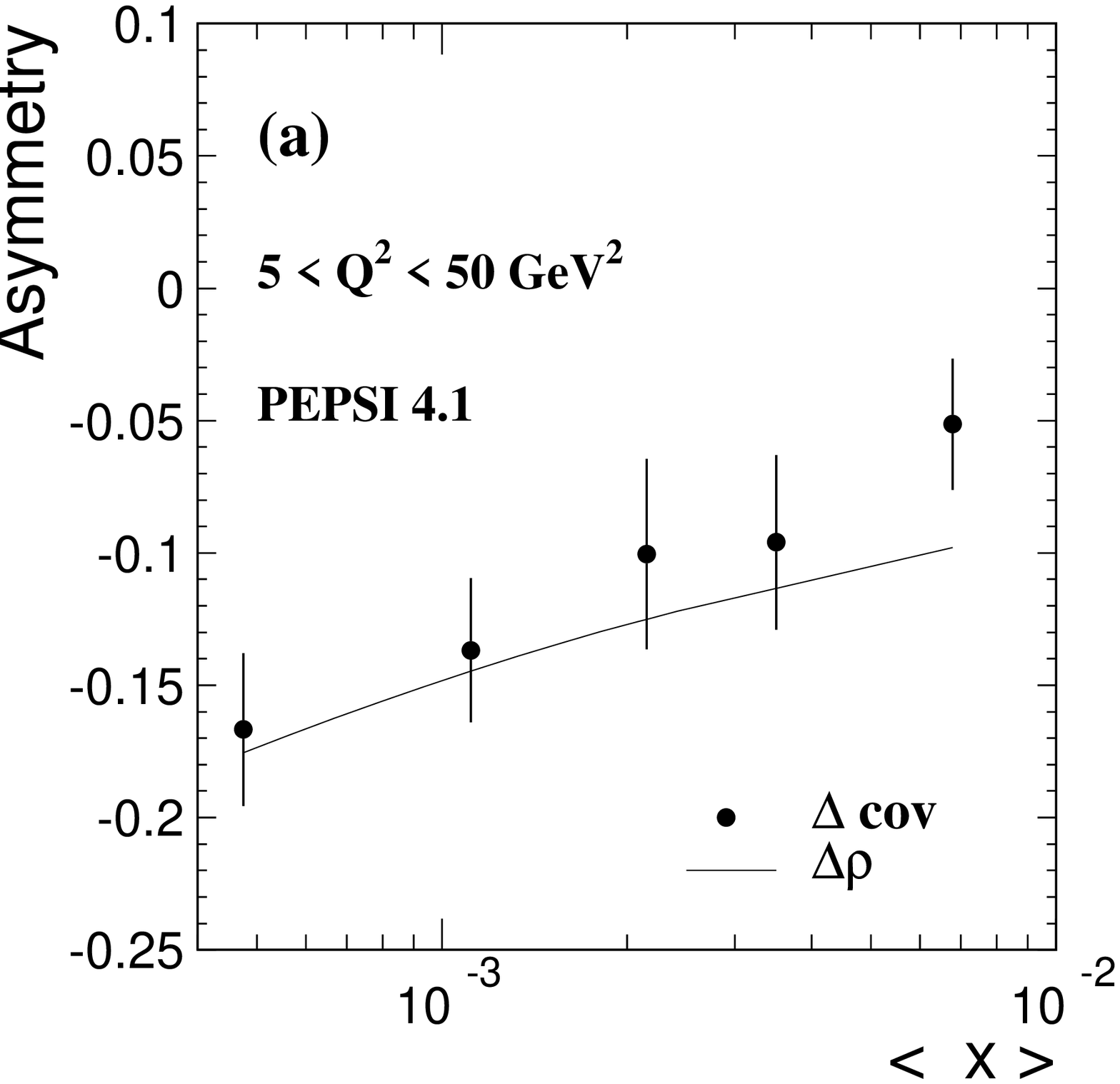,height=8.0cm}}
\put(95,0){\epsfig{file=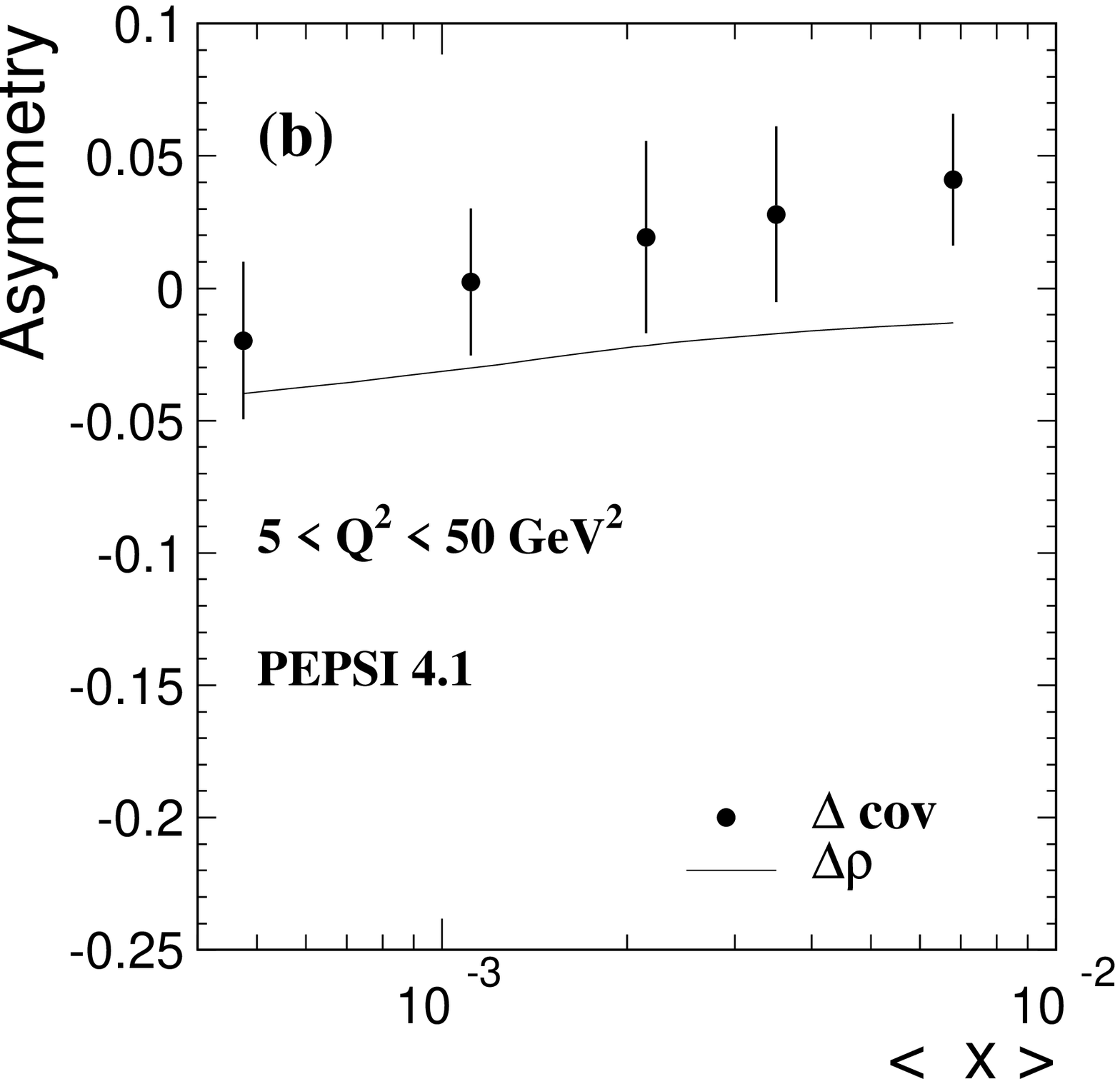,height=8.0cm}}
\end{picture}
\caption{{\it
The asymmetry calculated using  correlations
between current- and target-region charged multiplicities
generated with PEPSI (full symbols)
compared to the theoretical  expectations
for $\Delta\rho$ of the same  model (solid lines).
The figures show the $x$-dependence of the  asymmetry
in the range $5< Q^2 <50$ GeV$^2$: 
{\bf (a)} The asymmetry for the GS-A model implemented in PEPSI;
{\bf (b)} The same but when $\Delta G$ is set to zero.
}}
\label{fig2}
\end{center}
\end{figure}
 
Fig.~\ref{fig2}a shows 
the behavior of $\Delta\mathrm{cov}$ calculated
from final-state charged hadrons compared to   
the theoretical prediction for $\Delta\rho$, 
generated also with the PEPSI
using information on the generated type of the
LO process. 
The GS-A \cite{GS-A} polarized parton distribution was used.  
From this figure one can
conclude that the experimentally measured asymmetry 
$\Delta\mathrm{cov}$
does  reproduce the theoretically expected behavior of the
$\Delta\rho$. This is especially evident for
a small $x$, where the contribution from the QCDC
can be neglected.     
It is also seen that typical statistics expected at
HERA are sufficient to reliably determine the sign and the magnitude of the
asymmetry.

For an illustration, Fig.~\ref{fig2}b shows 
the same as Fig.~\ref{fig2}a, but 
after switching the gluon polarization term  off, i.e. $\Delta G=0$.
The quark polarization distributions were unchanged. 
In this case, both $\Delta\rho$ 
and $\Delta\mathrm{cov}$  tend towards zero. A 
difference between $\Delta\rho$ and zero value is expected to 
come from a positive value of $A_{||}$ (\ref{14}). 
It may also be noted a small discrepancy between
$\Delta\mathrm{cov}$ and $\Delta\rho$. This might mean that
for this, in fact, unrealistic case, the values of
$R_\bgf^{\an}$ and $R_\bgf^{\pa}$ are so close to each other
that an approximate nature of the expressions $(\ref{15})$ makes
a noticeable effect on the relationship between $\Delta\mathrm{cov}$
and $\Delta\rho$.

\bigskip
{\em Dependence on cut-offs.} 
The LEPTO Monte Carlo used by PEPSI contains cut-offs to
prevent divergences in the LO  matrix elements.   
Therefore,  the magnitude of 
the BGF rate in PEPSI is ambiguously defined.
This may produce a systematic bias in the relation (\ref{16}).

Let us remind that,  
in the  BGF cross section,  the singularities in the two-parton emission
are given by $1/z(1-z)$ \cite{mirk}, where    
$z=(p'\cdot p) / (p'\cdot q)$ 
($p$ is the momentum of the incoming parton and $p'$
is that of the final-state parton, $q^2=-Q^2$).
The LEPTO cut-offs  are based on the so-called ''mixed scheme`` in which
the parameter $z_{min}$,
restricting the values of the variable $z$, 
plays an important role in the determination
of the probability for the BGF event to occur. 
The default value of this cut-off is set to 0.04. 
If one decreases this cut-off, the relative contribution
of the BGF rates  increases, with respect to the QCDC.
In this case, the relation (\ref{7}) is expected to be a 
good estimate. However, for a large  $z_{min}$, 
the BGF rate is small and $\Delta \mathrm{cov}$ might poorly
reflect the behavior of $\Delta\rho$.   
Therefore, for our systematic checks, the
value of $z_{min}$ was increased. We have verified that the
relationship between $\Delta \mathrm{cov}$ and 
$\Delta \rho$ still holds up to $z_{min}=0.2$ for the given
statistic uncertainties. Note that while for the default
value $z_{min}=0.04$ the BGF rate is about $0.6$ in  the
smallest $x$-bin, the rate of this process 
drops down to $0.3$ when $z_{min}=0.2$ is considered.       

We have also found that $\Delta\rho$ changes only a little
by varying the cut-off. This can be explained by the  
normalisation $(R_\bgf^{\an} + R_\bgf^{\pa})$ used in (\ref{13}).  
Note that, from the experimental point of view, a similar
normalisation $(\mathrm{cov}^{\an} + \mathrm{cov}^{\pa})$
in  (\ref{16})  would help to measure the  
$\Delta\mathrm{cov}$ reliably even if the 
detector track acceptance for the
target-region is small (see \cite{ZEUS} for details).    

\bigskip
{\em Dependence on the structure function}.  
An important test for our method is to investigate  the asymmetry measured
with $\Delta \mathrm{cov}$ for different types of the 
parameterization for the parton densities. Various types of polarized parton 
distributions included into the PEPSI package were analyzed.     
The method was found to be sensitive to input structure functions 
and $\Delta \mathrm{cov}$ reproduces  the trends of $\Delta\rho$. 
 
\section{Conclusions}

In this paper we have proposed 
a new method to measure the asymmetry in 
polarized lepton-nucleon scattering. 
This method is based on the measurement of the current-target
correlations in the Breit frame. The advantage of the
method is that all inclusive DIS events with
well reconstructed Breit frame  can be used to determine the 
asymmetry, without specific constraints to select useful events 
for this measurement.  

In this respect, it is important to emphasize that our 
approach is well suited for rather low $Q^2$ and   
$E_T$, i.e. for the regions where dijet reconstruction 
suffers from misclusterings and large hadronization corrections.  
Thus the suggested  method  compliments and extends the study of $\Delta G/G$
to kinematic regions of low $Q^2$ ($E_T$) 
where the method suggested
in \cite{roeck} is less reliable.  
We also expect different systematics for these two methods: While the method 
discussed in \cite{roeck} is a subject of the systematic uncertainties    
as for the standard dijet studies, systematic effects for our method
most probably would come from a low detector acceptance in 
the target region of the Breit frame.

\section*{Acknowledgments}

We thank N.~Akopov, M.~Amarian and E.~Kuraev  
for helpful  discussions.

{}

\end{document}